\documentclass[prl,twocolumn,showpacs,floatfix]{revtex4}
\usepackage{graphicx,amsmath,amssymb,epsfig,color,multirow,dcolumn,bm}

\begin{document}

\title{Spin-Transport in Defective Graphene Nanoribbons}

\author{S.M.-M. Dubois}
\email[Electronic address:~]{Simon.Dubois@uclouvain.be}
\author{G.-M. Rignanese}
\author{J.-C. Charlier}
\affiliation{Universit\'e catholique de Louvain, \\
Unit\'e de Physico-Chimie et de Physique des Mat\'eriaux (PCPM), \\
European Theoretical Spectroscopy Facility (ETSF), \\
Place Croix du Sud 1, 1348 Louvain-la-Neuve, Belgium}

\date{\today}

\begin{abstract}

Using first-principles calculations, the effect of magnetic point defects
(vacancy and adatom) is investigated in zigzag graphene nanoribbons.  The
structural, electronic, and spin-transport properties are studied.  While
pristine ribbons display anti-parallel spin states at their edges, the
defects are found to perturb this coupling. The introduction of a vacancy
drastically reduces the energy difference between parallel and anti-parallel
spin orientations, though the latter is still favored. Moreover, the local
magnetic moment of the defect is screened by the edges so that the total
magnetic moment is quite small.  In contrast, when an adatom is introduced,
the parallel spin orientation is preferred and the local magnetic moment of
the defect adds up to the contributions of the edges. Furthermore, a
spin-polarized transmission is observed at the Fermi energy, suggesting the
use of such a defective graphene nanoribbon as spin-valve device.

\end{abstract}


\maketitle

Recent progresses in the preparation of contacted graphene mono-layer have
enabled the synthesis of well-controlled nanometer-sized systems with open
boundaries~\cite{novoselov1,berger1}.  Both one-dimensional graphene
ribbons and zero-dimensional graphene dots have been achieved either by
cutting exfoliated graphene layers~\cite{bunch1,geim1} or by patterning
epitaxially grown graphenes~\cite{han1}. Chemical methods have 
also been designed to achieve a solution-phase derivation of graphene 
ribbons with ultrasmooth edges~\cite{li1}.  These advances are at the origin
of the raising interest in open boundary systems where the presence and shape
of the edges influence drastically the $\pi$-electronic structure~\cite{koskinen}, 
opening the road towards new physical phenomena and technological
applications. 
 
As far as ribbons are concerned, the geometry of $sp^2$-bonded networks
implies two possible cutting directions called ``zigzag" and ``armchair"
according to the shape of the created edge. Due to topological reasons,
zigzag-shaped edges give rise to peculiar extended electronic states which
decay exponentially inside the graphene sheet~\cite{fujita1,nakada1}.
These edge-states, which are not reported along the armchair-shaped edges,
come with a twofold degenerate flat band at the Fermi energy over one third
of the Brillouin zone and have been suggested to be ferromagnetically ordered
along the edge~\cite{lee1}. Recent DFT calculations on GNRs with
zigzag-shaped edges (zGNRs), have suggested that due to the finite width of
the ribbon, the interaction between the edges favors an opposite spin
orientation at these edges~\cite{lee1,lee3,son1}.  These predictions are
in agreement with the second Lieb's theorem~\cite{lieb1} about the total
spin in bipartite lattices.

Besides the possible magnetization of the edges, defect-induced magnetism is
also expected in GNRs. Indeed, while many common topological defects (5/7
pairs, Stone-Wales~\cite{stone1},...) do preserve the saturation of the
carbon atoms and are reported not to cause any local magnetization~\cite{duplock1}, 
processes where dangling bonds are created, may possibly induce a
local magnetic moment~\cite{lehtinen1,lehtinen2,yazyev1}.  For instance,
carbon atoms can be removed from their lattice position, creating pairs of
vacancies and adatoms.  Such Frenkel defects are frequently present at
thermal equilibrium~\cite{ebbesen1,kosaka1} in $sp^2$-bonded carbon
systems.  Depending on their positions, such magnetic point defects might
affect differently the magnetic structure of GNRs. When located at the edge,
these defects affect mainly the ferromagnetic order along the edge~\cite{wimmer1}.  
However, when located inside the ribbon, their effect on the
magnetic structure is still an open issue.

In the present letter, magnetic point defects located in the vicinity of the
ribbon axis have been used to tailor both the electronic and spin-transport
properties of hydrogen passivated zGNRs. More specifically, the introduction
of vacancy/adatom in these systems decreases the energy difference between
the parallel and anti-parallel spin orientations of the edges. The magnetic
moment of the vacancy defect is found to be drastically reduced in the
defective ribbon compared to the ideal one. In the presence of carbon
adatoms, a parallel spin orientation at the edges is favored, giving rise to
a spin-polarized electronic transmission function around the Fermi energy. 

Electronic irradiations can induce such magnetic point defects in graphene
samples and allow to tune their number and their concentration~\cite{esquinazi1}.  
Therefore, based on our theoretical predictions, one can hope
to use defective zGNRs as spin-based nano-electronic devices.

Spin Polarized Density Functional Theory~\cite{soler1,artacho1} within the
generalized-gradient approximation is used to investigate a hydrogen
passivated 8-zGNR (width $\sim18$\AA).  Periodic boundary conditions are used
with fixed lateral dimensions which insure 25~\AA~of vacuum between the GNRs
in neighboring cells.  When vacancies or carbon adatoms are introduced, a
6$\times$1$\times$1 supercell is considered, leading to an
nearest-neighboring defect distance of $\sim15$~\AA\ and a defect
concentration of $\sim1$\%.  Note that, for the computation of the
transmission functions, a 22$\times$1$\times$1 supercell is considered in
order to obtain a good screening of the perturbed Hartree potential due to
the defect.  In order to deal with the large number of atoms in the
supercells,  numerical atomic orbital basis set~\cite{basis} are used to
expand the wave-functions in conjunction with norm-conserving
pseudo-potentials. The energy levels are populated using a Fermi-Dirac
distribution with an electronic temperature of 250K.  The integration over
the 1D Brillouin zone is replaced by a summation over a regular grid of 40
\textit{k}-points~\cite{note_kpt}.  The geometry is fully relaxed until the
forces on each atom, and on the unit cell~\cite{note_stress} are less than
0.01 eV/\AA\ and 0.05 eV/\AA\ respectively.  The small spin-orbit coupling of
carbon atoms is neglected and independent collinear spin orientations,
labeled $\alpha$ and $\beta$, are considered.  Within such framework, the
electronic ground state of the ideal 8-zGNR reveals anti-parallel ($\uparrow
\downarrow$) spin orientations between the edges leading to a semiconducting
band structure (0.5 eV band gap) with full spin degeneracy.  On the other
hand, the magnetic configuration with parallel ($\uparrow \uparrow$) spin
orientations between the edges is found to be metastable (11~meV/edge-atom
higher in energy). This configuration displays a metallic behavior, as the
$\pi^{*}_{\alpha}$ and $\pi_{\beta}$ bands are crossing at the Fermi energy,
and presents a total magnetic moment of 0.51~$\mu_B$ per edge atom.  These
results are in good agreement with the previous calculations~\cite{lee1,son1}.

By analogy with graphene, the removal of a carbon atom (vacancy creation)
induces a Jahn-Teller distortion of the honeycomb structure. Two of the
unsaturated carbon atoms come closer one to the other and form a weak
covalent bond, inducing a pentagon like rearrangement.  The third unsaturated
carbon atom moves out of the plane~\cite{lehtinen1,barbary1}.
Consequently, the initial $D_{3h}$ symmetry of the hexagonal network is lost
in favor to the $C_s$ symmetry~\cite{amara1}. In zGNRs, such atomic
rearrangement leads to two possible orientations of the vacancy, called
``perpendicular" and ``tilted" according to the relative orientation of the
mirror plane of the vacancy with respect to the ribbon axis.

\begin{figure}[tb]
\includegraphics{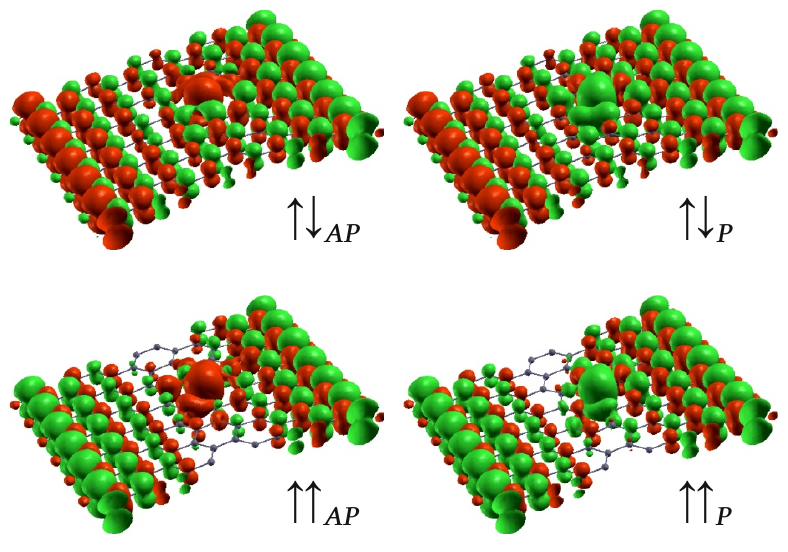}
\caption{[Color online] Spin polarized electronic densities ($\rho_{\alpha} -
\rho_{\beta}$) isosurfaces for the four magnetic configurations associated to
the tilted vacancy in zGNR. Green and yellow surfaces correspond to an excess
of spin $\alpha$ and spin $\beta$ electrons, respectively.}
\label{hole}
\end{figure}

As expected due to the presence of dangling bonds, the electronic density
appears to be locally spin-polarized on the vacancy. The interaction between
the magnetic moment of the defect and the spin ordered edge states results in
four locally stable configurations for both vacancy orientations (see
Fig.~\ref{hole}).  In the following, the various magnetic configurations will
be labeled $X_Y$, where $X$ refers to the parallel ($\uparrow \uparrow$) or
anti-parallel ($\uparrow \downarrow$) spin orientations between the ribbon
edges, while $Y$ refers to the parallel ($P$) or anti-parallel ($AP$) spin
orientations between the defect and the nearest edge. 

\begin{table}[tb]
  \begin{tabular}{lrrlrr}
    \hline
     & \multicolumn{2}{c}{Tilted vacancy} & & \multicolumn{2}{c}{Perpendicular vacancy}     \\
     & $\Delta E$ (eV) & $M_{tot}$ ($\mu_B$) & & $\Delta E$ (eV) & $M_{tot}$ ($\mu_B$) \\
    \cline{2-3} \cline{5-6}
     $ \uparrow \downarrow_{AP}$  & 0.000 & 0.19 & & 0.110  & 1.98 \\ 
     $ \uparrow \downarrow_{P}$   & 0.004 & 1.68 & & 0.142  & 0.06 \\ 
     $ \uparrow \uparrow_{AP}$   & 0.021 & 2.03 & & 0.154  & 2.10 \\ 
     $ \uparrow \uparrow_{P}$    & 0.023 & 4.06 & & 0.170  & 4.52 \\ 
    \hline
  \end{tabular}
\caption{Energies related to the ground state ($\uparrow \downarrow_{AP}$)
and total magnetic moments associated to the eight possible magnetic
configurations of the vacancy.}
\label{vacancy1}
\end{table}

The energies related to the eight possible magnetic configurations of the
vacancy are detailed in Table~\ref{vacancy1}. The tilted vacancy is found to
be the most stable orientation, probably because of the formation of a
stronger $C-C$ bond during the atomic reconstruction process. Consequently,
the ribbon width displays a small shrinking (1.1\%) around the vacancy, while
a stretching is observed for the perpendicular vacancy orientation. Both
vacancy orientations have the same $\uparrow \downarrow$ ground state as the
pristine GNR.  However, the energetics of the inter-edge coupling is
modified, as shown in Table~\ref{vacancy1}. Indeed, while the energy
difference between the parallel ($\uparrow \uparrow$) and anti-parallel
($\uparrow \downarrow$) spin orientations at the edges is 66~meV in the
absence of defect, this value is reduced to 21~meV (44~meV) in the presence
of a tilted (perpendicular) vacancy. At last, an accurate convergence study
with respect to the size of the supercell allows us to predict a formation
energy of $15.6$~eV (which is approximately twice the one reported for the
single vacancy in the graphene sheet~\cite{lehtinen1,yazyev1}) and a weak
magnetic moment of 0.15~$\mu_B$ for the isolated vacancy in the tilted
$\uparrow \downarrow_{AP}$ configuration.

\begin{figure}[tb]
\includegraphics{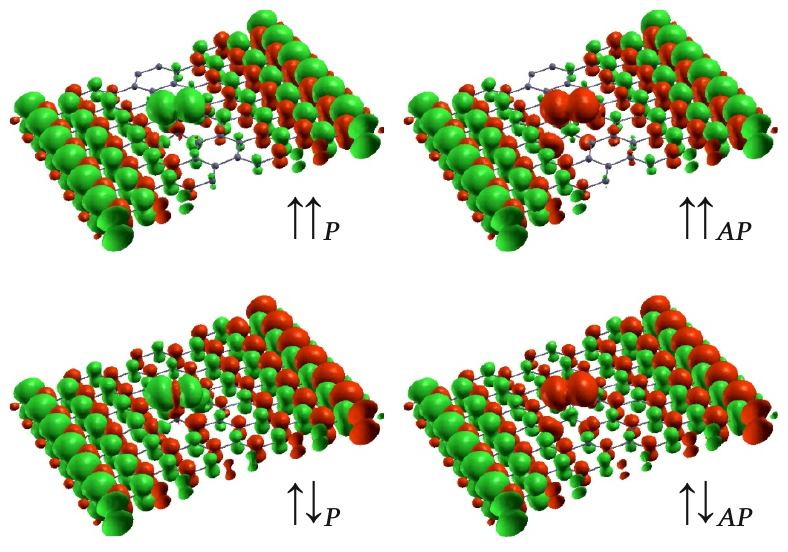}
\caption{[Color online] Spin polarized electronic densities ($\rho_{\alpha} -
\rho_{\beta}$) isosurfaces for the four magnetic configurations associated to
the carbon adatom in the tilted orientation. Green and yellow surfaces
correspond to an excess of spin $\alpha$ and spin $\beta$ electrons,
respectively.}
\label{adatom}
\end{figure}

At equilibrium, the carbon adatom is in a bridge-like position between two
in-plane carbon atoms. This geometry is similar to the one reported for
carbon atoms on graphene~\cite{lehtinen1,lehtinen2,nordlund1,lee2,amara1}. 
Again, in zGNRs, two adsorption sites are possible: labeled
perpendicular or tilted depending on the orientation of the underlying
carbon-carbon bond with respect to the ribbon axis. In both cases, the carbon
adatom displays a local magnetic moment.  In analogy with the vacancy case,
the various magnetic configurations are labeled using the aforementioned
$X_Y$ syntax (see Fig.~\ref{adatom}), and their corresponding energies are
detailed in Table~\ref{adatom1}.

\begin{table}[tb]
  \begin{tabular}{lrrllrr}
    \hline
     & \multicolumn{2}{c}{Tilted adatom}  & & & \multicolumn{2}{c}{Perpendicular adatom}    \\
     & $\Delta E$ (eV) & $M_{tot}$ ($\mu_B$) & & &$\Delta E$ (eV) & $M_{tot}$ ($\mu_B$) \\
    \cline{2-3} \cline{6-7}
     $\uparrow \uparrow_{P}$    &  0.000 & 3.73 & &  &  0.060 & 4.00     \\ 
     $\uparrow \uparrow_{AP}$   &  0.005 & 2.58 & &  &  0.089 & 2.81     \\
     $\uparrow \downarrow_{P}$   &  0.008 & 0.11 & & &  0.094 & 0.72   \\ 
     $\uparrow \downarrow_{AP}$  &  0.032 & 0.67 & &  &  0.094 & 0.72 \\
    \hline
  \end{tabular}
\caption{Energies related to the ground state ($\uparrow \uparrow_{P}$) and
total magnetic moments associated to the eight possible magnetic
configurations of the carbon adatom.}
\label{adatom1}
\end{table}

The tilted $C-C$ bond is found to be the preferred host for carbon adsorption
as it is associated with a smaller in-plane $C-C$ bond and a larger distance
between the adatom and the ribbon. No significant shrinking is reported here.
Similar to vacancies, carbon adatoms modify the energetics of the inter-edge
coupling.  However, this effect is enhanced in the presence of carbon adatoms
and the $\uparrow \uparrow$ ground state is different from the one the
pristine GNRs.  Finally, an adsorption energy of $1.2$~eV and a magnetic
moment of 0.48~$\mu_B$ are computed for the isolated carbon adatom in the
$\uparrow \uparrow_{P}$ configuration. These values are quite similar to the
$1.40$~eV and $0.45$$\mu_B$ reported for the adsorption of carbon on
graphene~\cite{lehtinen1}.

While the resulting ground states are different in the presence of a vacancy
($\uparrow \downarrow$) or a carbon adatom ($\uparrow \uparrow$), both
defects induce the same trend on the inter-edge coupling.  Indeed, the energy
difference between the $\uparrow \downarrow$ and the $\uparrow \uparrow$
configurations in the presence of defects is found to be reduced well below
the 66~meV predicted for the pristine ribbon.  This result is a direct
consequence of the defect-induced perturbation of the spin-polarized density.
On one hand, by breaking the bipartite character of the carbon network,
defects partially destroy the enhanced exchange splitting responsible of the
stabilization of the $\uparrow \downarrow$ configuration~\cite{lee1}.  On the
other hand, as can be seen from Figs.~\ref{hole} and~\ref{adatom} the spin
polarization is larger in the center of the ribbon for the $\uparrow
\downarrow$ configuration than for the $\uparrow \uparrow$ one. As a result,
the density rearrangement is more limited around the defects in the former,
which is confirmed by the projected density of states (not shown
here). In summary, the introduction of defects induces a stabilization of the
$\uparrow \uparrow$ configuration in contrast with the pristine case.

\begin{figure}[htb]
\includegraphics{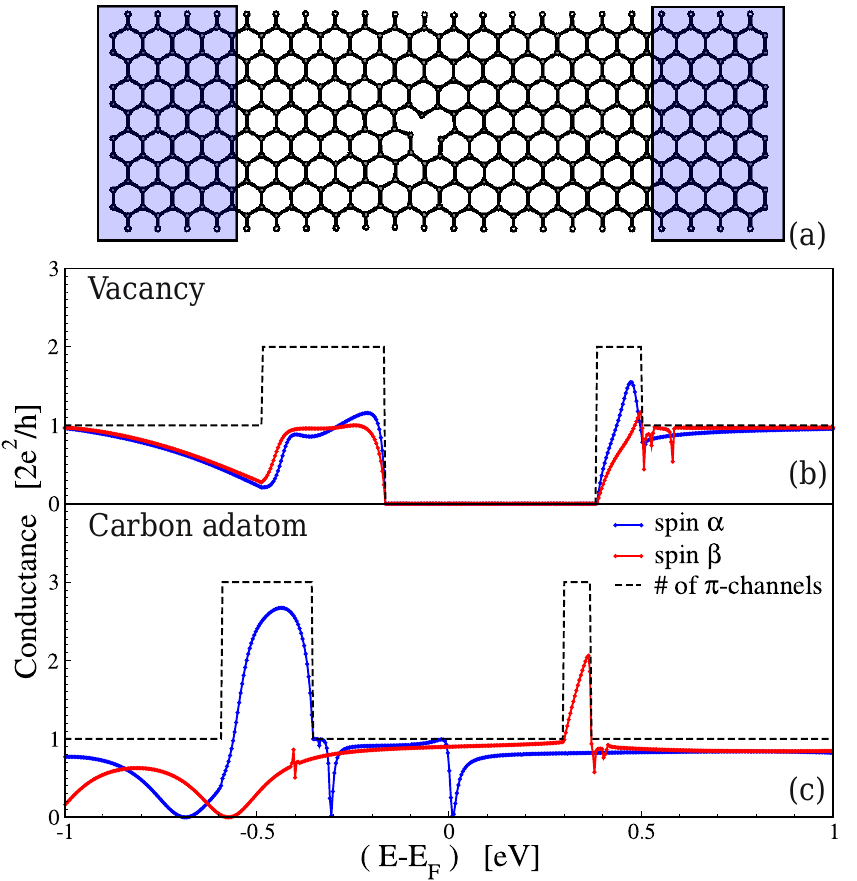}
\caption{[Color online] (a) Schematic representation of the model structure
used to predict the electronic transmission functions of the zGNR in the
presence of (b) an isolated vacancy in the $\uparrow \downarrow_{AP}$
configuration, and (c) an isolated carbon adatom in the $\uparrow
\uparrow_{P}$ configuration. The spin $\alpha$ and spin $\beta$ components
are appear in blue and red, respectively, while the dotted line indicates the
number of $\pi$ channels.}
\label{transm}
\end{figure}

Since the introduction of magnetic point defects in zGNRs favors a specific
spin configuration of the edges, a major impact is also expected on the
transport properties of these 1D systems~\cite{kim1}. This will be
investigated in the remainder of this letter. Our \textit{ab initio}
calculations of the electronic transmission functions, reported in
Fig.~\ref{transm}, are performed within the Landauer approach.  In order to
simulate open boundary conditions, self-energies associated with the leads
are included in the self-consistent calculation of the potential~\cite{rocha1}.  
The supercell containing the point defect is connected to two leads
consisting of a few unit cells of ideal zGNR [Fig.~\ref{transm}(a)].  Note
that, as indicated above, a larger supercell has been used in order to ensure
an accurate alignment of the electronic levels of the lead with those of the
supercell.

As mentioned before, the $\uparrow \downarrow$ spin configuration of the
pristine zGNR is preserved when an isolated vacancy is introduced.  The
defective zGNR remains semiconducting and its electronic transmission
function displays a gap of 0.5 eV around the Fermi energy.  On the contrary,
in the presence of a carbon adatom, the $\uparrow \uparrow$ spin
configuration is favored and the zGNR becomes metallic, inducing a non-zero
electronic transmission function at the Fermi energy. 

The main impact of the magnetic point defects on the transport properties is
a global reduction of the transmission associated with the $\pi$ and
$\pi^{*}$ electrons. This can be related to a decrease of the transmission
probability of some $\pi$-$\pi^{*}$ conduction eigenchannels compared to the
pristine zGNR.  For the vacancy [Fig.~\ref{transm}(b)], this effect appears
essentially for energies ranging from -0.80 to -0.15 eV ($\pi$ channels) and
from 0.4 to 0.5 eV ($\pi^{*}$ channels), inducing a slight breaking of the
spin degeneracy.  For the carbon adatom, a similar reduction of the
transmission is found.  But, defect-induced drops also appear at -0.3 eV and
0.02 eV. At these energies, states localized on the defect are present, which
can only mix with the spin $\alpha$ conduction channel.  Consequently, its
transmission probability goes to zero due to the alteration of its $\pi^{*}$
character and the spin-degeneracy of the electronic transmission function is
lifted up just above the Fermi energy.

In conclusion, the $\uparrow \uparrow$ spin configuration of zGNRs tends to
be favored by the introduction of magnetic point-defects in the 1D system, as
suggested by the low-energy spin-polarized density.  The major impact of
these point defects on the transport properties of zGNRs has been
investigated. The predicted spin-polarized transmission function underlines
the possibility to use adatom-doped graphene nanoribbons as valuable
nano-devices in future spin-based electronics. 
 
The authors acknowledge financial support from the F.R.I.A. (SMMD) and the
F.R.S.-FNRS (JCC and GMR).  Parts of this work are directly connected to the
Belgian Program on Interuniversity Attraction Poles (PAI6) on ``Quantum
Effects in Clusters and Nanowires'', to the ARC sponsored by the Communaut\'e
Fran\c{c}aise de Belgique and to the NANOQUANTA European Network of
Excellence. Computational resources have been provided by the Universit\'e  
catholique de Louvain~: all the numerical simulations have been performed on the
LEMAITRE computer of the CISM.


\begin{thebibliography}{99}
\bibitem{basis}
A numerical atomic orbital basis set of double-$\zeta$ plus one polarization
and the 3s orbitals (DZP+3S) quality has been used for all the structural
optimizations while the transmission functions have been obtained using basis
set of single-$\zeta$ plus one polarization and the 3s orbitals (SZP+3S)
quality.
\bibitem{note_stress}
The force on the unit cell along the GNR axis is obtained by multiplying the
corresponding stress by the surface of the unit cell in that direction.
\end{thebibliography}
\end{document}